\documentclass[%
 reprint,
 amsmath,amssymb,
 aps,
]{revtex4-2}

\usepackage{graphicx}
\usepackage{dcolumn}
\usepackage{bm}
\usepackage{array}
\usepackage[all,dvips]{xy}
\usepackage{epsfig}

\usepackage{epstopdf}

\begin{document}

\preprint{APS/123-QED}

\title{Chimera states for directed networks}

\author{Patrycja Jaros$^1$, Roman Levchenko$^{2,3}$, Tomasz Kapitaniak$^1$, Yuri Maistrenko$^{1,3,4}$}

\affiliation{$^{1}$Division of Dynamics, Lodz University of Technology, Stefanowskiego 1/15, 90-924 Lodz, Poland}

\affiliation{$^{2}$Forschungszentrum J\"{u}lich, 52428 J\"{u}lich, Germany}

\affiliation{$^{3}$National University of Kyiv, Volodymyrska St. 60, 01030 Kyiv, Ukraine}

\affiliation{$^{4}$Institute of Mathematics and Centre for Medical and Biotechnical Research, National Academy of Sciences of Ukraine, Tereshchenkivska St. 3, 01030, Kyiv, Ukraine}

\begin{abstract}

We demonstrate that chimera behavior can be observed in ensembles of phase oscillators with unidirectional coupling. For a small network consisting of only three  identical oscillators (cyclic triple), tiny {\it chimera islands} arise in the parameter space. They are surrounded by developed chaotic switching behavior caused by a collision of rotating waves propagating in opposite directions. For larger networks, as we show for hundred oscillators (cyclic century), the islands merge into a single {\it chimera continent} which incorporates the world of chimeras of different configuration. The phenomenon inherits from networks with intermediate ranges of the unidirectional coupling and it diminishes as it decreases.
  
 \end{abstract}
\maketitle

{\bf Reflecting the unrivaled harmony of the world, nonlinear oscillatory networks keep astonishing us with new types of fascinating collective behaviors. Chimera states have opened a new line of research underlining a typicality for the coexistence of synchrony and asynchrony as a result of the symmetry-breaking transition. Chimeras normally arise in networks with symmetric bi-directional coupling, however, start traveling and disappear under relatively slight violation of the coupling symmetry. In the present study, we demonstrate that, counter-intuitively, chimera states can also exist in networks with unidirectional, one-sided coupling. They arise in  the form of regular standing waves, playing the role of stoppers for the chaotic switching behavior between  clock- and anti-clock wise rotating waves.}

\section{INTRODUCTION}

Chimera states represent a fast growing field of nonlinear science, reflecting a possibility for symmetric system to behave in a hybrid way such that regular and irregular behavior co-exist. The state of art of the chimera study are well presented in recent review papers, see  \cite{pa2015, s2016, omel2019, anna2020, chimeras}, which demonstrates an amazing variety of the chimera types and nonlinear systems where they are obtained. The robustness of the chimera phenomenon is confirmed by many experimental observations, which give hope that chimera states may play a role in the understanding of peculiar complex behaviors in biological \cite{network, birds, epilepsy,  neuro}, engineering \cite{grid,pshmr2014, squids, belykh2017, delayed, powergrid}, and social \cite{social, pik2021} systems.

Beginning from the pioneering works \cite{ref2, ref3}, chimera states are mainly reported for networks of oscillators with symmetric, both-sided (if on a ring) coupling. In the classical Kuramoto-Sakaguchi model, they are characterized by chaotic wandering in the space \cite{drift2010},  which makes them extremely sensitive to any level of the coupling asymmetry.  First, the chimeras start traveling with a constant velocity (proportional to the asymmetry), and then   disappear in a crisis  at a relatively small level of  asymmetry \cite{asym2013, xie2014, bick2015}. 

Such  "fragile" behavior of the chimeras in classical Kuramoto model is overcome by the accounting of inertia. Although the chimera chaotic wandering becomes even more pronounced (identified as ''imperfect chimera state'' \cite{kkwcm2014}), there arises a new type of the chimeric behavior, called {\it solitary states} \cite{jar2015,jar2018}. They represent an essential sub-class of the weak chimera states \cite{ab2015}, in which only one or a few oscillators split up from the main synchronized cluster and start to rotate with different average frequency (Poincar\'e winding number). Solitary states appear to be qualitatively different from the classical chimeras discovered originally in \cite{ref2, ref3}. They do not wander in space and mathematically, are Lyapunov stable states (standing waves) in the system phase space. Recently, the existence of the solitary states has been also reported in small networks \cite{jar2021}, adaptive \cite{adaptive}, multiplex \cite{anna2021}, and  power grid \cite{powergrid} systems,  as well as in the mean-field limit \cite{kruk2020}. 

An exclusive characteristic of the solitary states is their strong robustness with respect to the violation of the coupling symmetry (in contrary to the classical chimeras in standard Kuramoto model).  In this study, we report the solitary state appearance for a maximally asymmetric, unidirectional network of $N$ oscillators in a ring, coupled in such a way that each node is under action of some number $P<N/2$ of its neighbors from one side, but not from the other.

\section{COUPLED TRIPLE: N=3 CASE.}

The network dynamics is described by a modified Kuramoto model with inertia \cite{onbt2014, jar2015, jar2018,jar2021} of the form:

\begin{equation}
m\ddot{\theta_{i}}+\varepsilon\dot{\theta_{i}}=\omega+\dfrac{\mu}{P}\sum_{j=1}^{P}\sin(\theta_{i+j}-\theta_{i}-\alpha), \label{eq:} 
\end{equation}
\noindent
where $\theta_i(t), i=1,...N$, are phase variables, $\alpha$ is a phase lag, and  $\mu$ is a coupling strength. Other parameters:  $m$, $\varepsilon$, and $\omega$ are mass, damping, and natural frequency of a single oscillator. Without loss of generality, we will put $m=1.0$, $\varepsilon=0.1$, and $\omega=0$ and explore the complex dynamics of the model (1) varying parameters $\mu>0$ and $0<\alpha<\pi/2$. To ensure a directed coupling between any two oscillators, radius $P$ is assumed smaller than $N/2$. (For a definiteness,  each oscillator in (1) is under the influence of the neighbors from the right; equivalently, the left-side coupling topology will gives the same results). Regarding the terminology, we use below the term "chimera state" as more general, instead of "solitary state".  

Consider first the smallest non-trivial network of only $N=3$ oscillators with $P=1$, i.e., of the so-called {\it cycling triple} graph configuration.
Eq.(1) is then a 6-dimensional system of differential equations.   Its effective dynamics is, however,  a 4-dimensional given by the reduced system in phase differences $\eta_1=\theta_1-\theta_2$, $\eta_2=\theta_1-\theta_3$:

\begin{equation}
\begin{split}
m\ddot{\eta_1}+\varepsilon\dot{\eta_1}& =-\sin(\eta_1+\alpha)+\sin(\eta_2+\alpha),  \\ 
m\ddot{\eta_2}+\varepsilon\dot{\eta_2}& =-\sin(\eta_2+\alpha)-\sin(\eta_1+\eta_2-\alpha). 
\label{KurInertia2}
\end{split}
\end{equation}

Reduced system (2) has three equilibria:  ($\eta_1, \eta_2)=(0,0)$,  $(-2\pi/3, 2\pi/3)$, and $(2\pi/3, -2\pi/3)$ corresponding to three phase-locked states of the model (1). They are the fully synchronized state $O\equiv(\theta_1(t)=\theta_2(t)=\theta_3(t))$ and two splay states $S^{(\pm)}\equiv(\theta_i(t)=\pm2{\pi}i/3,  ~i=0,1,2$).  Stability of the states is controlled by the characteristic polynomial 
\begin{center}
	$\lambda^4+2\varepsilon\lambda^3+( \varepsilon^2+3a)\lambda^2+3a\varepsilon\lambda+3a^2=0,$
	\label{ch_equation}
\end{center}
where $a=\mu\cos\alpha$ for synchronous state $O$, and $a=\mu\cos(\alpha\pm2\pi/3)$ for splay states $S^{(\pm)}$. Roots of the characteristic polynomial (2) have negative real parts if and only if  parameter $a$ belongs to the interval $(0,2\varepsilon)$. It follows that synchronous state $O$  is stable in the $(\alpha,\mu)$-parameter region  $0<\mu<2\varepsilon^2/{\cos\alpha}$, $-\pi/2\alpha<\pi/2$; splay state $S^{(+)}$ in the region $0<\mu<2\varepsilon^2/{\cos(\alpha-2\pi/3)}$, $\pi/6<\alpha<7\pi/6$; and $S^{(-)}$ in the region $0<\mu<2\varepsilon^2/{\cos(\alpha+2\pi/3)}$, $-7\pi/6<\alpha<-\pi/6$.

In the study, we will analyze the system dynamics at the attractive parameter-$\alpha$ interval $[0,\pi/2)$, where only two of the three phase-locked states can stabilize: $O$ on the whole interval and splay state $S\equiv S^{(+)}$ for $\alpha>\pi/6$.  A peculiarity of the phase-locked dynamics is that $O$ and $S$ rotate in opposite directions: $O$ is characterized by a negative velocity $-(\mu/\varepsilon)\sin\alpha$, while $S$ by a positive velocity $(\mu/\varepsilon)\sin(2\pi/3-\alpha)$; the velocities coincide (in modulus)  at $\alpha=\pi/3$.
\begin{figure}
	\includegraphics[scale=1.0]{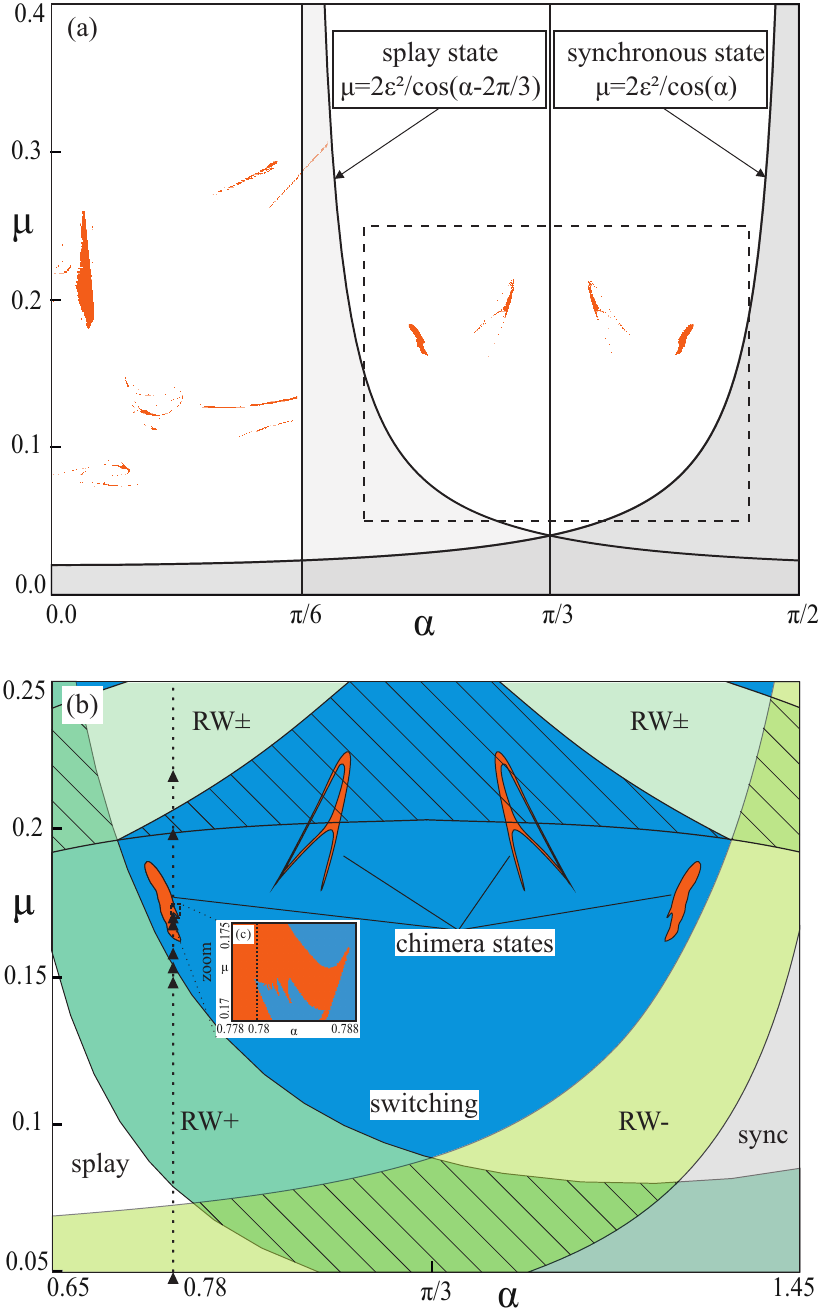}
	\caption{(color online) (a) Phase diagram for system (1) with
		N = 3 (cyclic triple coupling configuration) in the $(\alpha, \mu)$ parameter plane. Regions of chimera states are shown in orange, the synchronous and splay state regions are shaded in gray and light gray, respectively. (b) Enlargement of the delineated rectangle from (a): Blue region stands for chaotic switching behavior between the two saddle rotation waves $RW^{(+)}$ and $RW^{(-)}$ (see the text); the waves are stable in respective colored region below; $RW^{(\pm)}$ wave is stable above; the dynamics is multistable in hatched region. Stability regions for synchronous and splay states are left blank. Inset (c) shows an enlargement of a part of the left chimera island in (b).  Parameters $m=1.0$, $\varepsilon=0.1$.}
	\label{fig1}
\end{figure}

It has to be noted that due to the symmetry of model (1), its effective dynamics is actually encased at the $\alpha$-interval $[0,\pi/3]$. Indeed, if parameter $\alpha$ lies beyond this interval, Eq.(1) can always be reduced to an equivalent form with a symmetrically located values of  ${\alpha}\in[0,\pi/3]$,  by one of the following two transformations: 1) $\alpha\mapsto-\alpha$,  $\theta_i\mapsto-\theta_i$  or 2) $\alpha\mapsto2\pi/3-\alpha$,  $\theta_i\mapsto-\theta_i+2{\pi}i/3$ ($i=0,1,2$).
The first transformation is well known for general Kuramoto model - it allows to study the model only for positive $\alpha$. The second one is specific to the cycling triple coupling configuration of Eq.(1). It establishes  the system symmetry with respect to the $\alpha=\pi/3$ axis, and it maps the fully synchronous $O$ and the splay states $S^{(\pm)}$ to each other. Under the transformation, the main dynamical features of model (1) are left identical, as it is seen in the parameter bifurcation diagram in Fig.1.

\begin{figure}
	\includegraphics[scale=0.57]{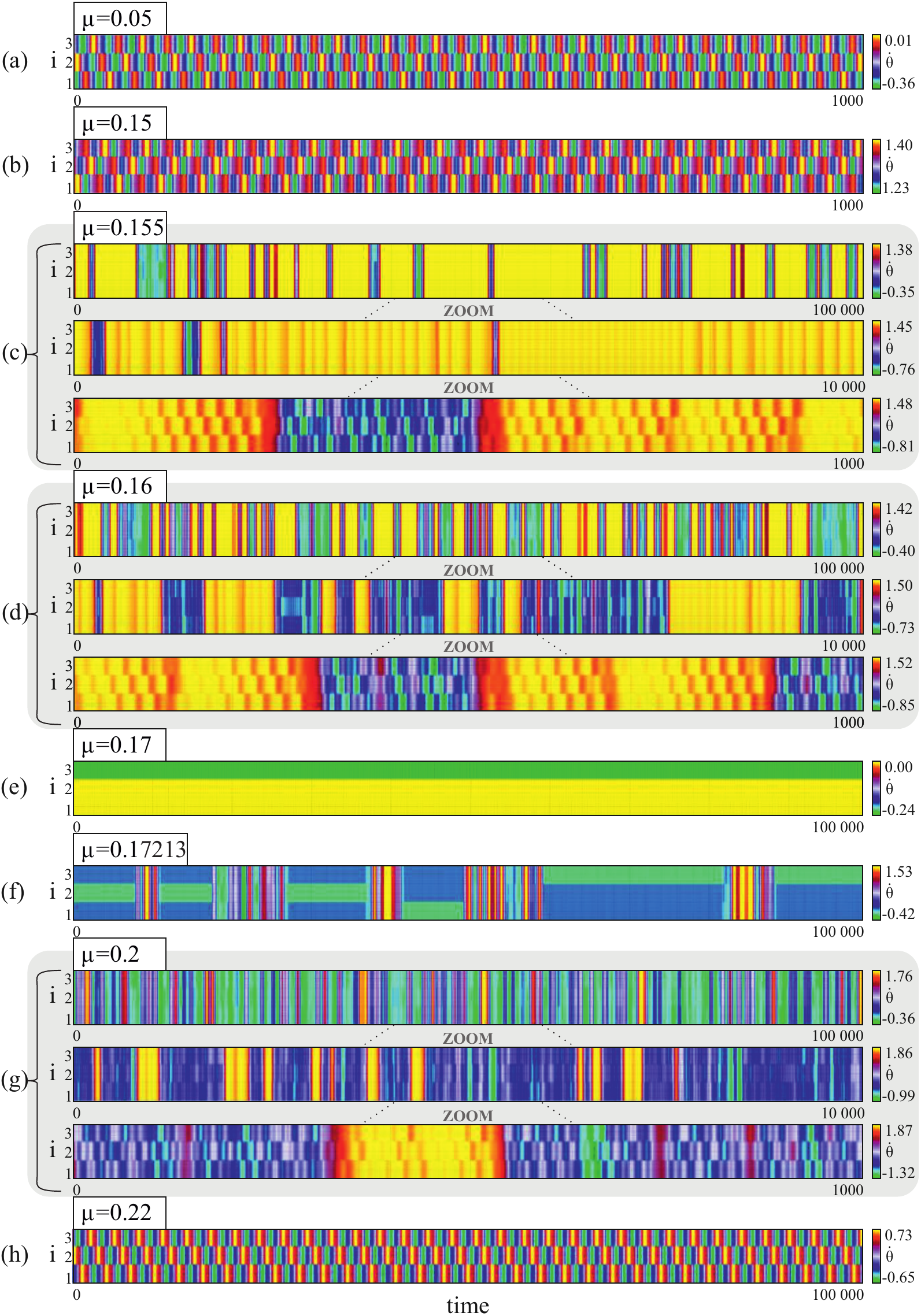}
	\caption{(color online) Coherence-incoherence chimera transition in system (1) with $N=3$. Frequency-time plots are shown for fixed phase lag $\alpha=0.78$,  and the coupling parameter $\mu$ increasing along the dashed vertical line with triangles in Fig.1(b), from top to down. (a,b,h): rotating waves $RW^{(-)}$, $RW^{(+)}$, and $RW^{(\pm)}$, respectively; (c,d,g): chaotic heteroclinic switching between $RW^{(-)}$ and $RW^{(+)}$; (e) chimera state; (f) heteroclinic switching between chimera states. Parameters $m=1.0, \varepsilon=0.1$. }
	\label{fig2}
\end{figure}

\begin{figure}[hb!]
	\includegraphics[scale=1]{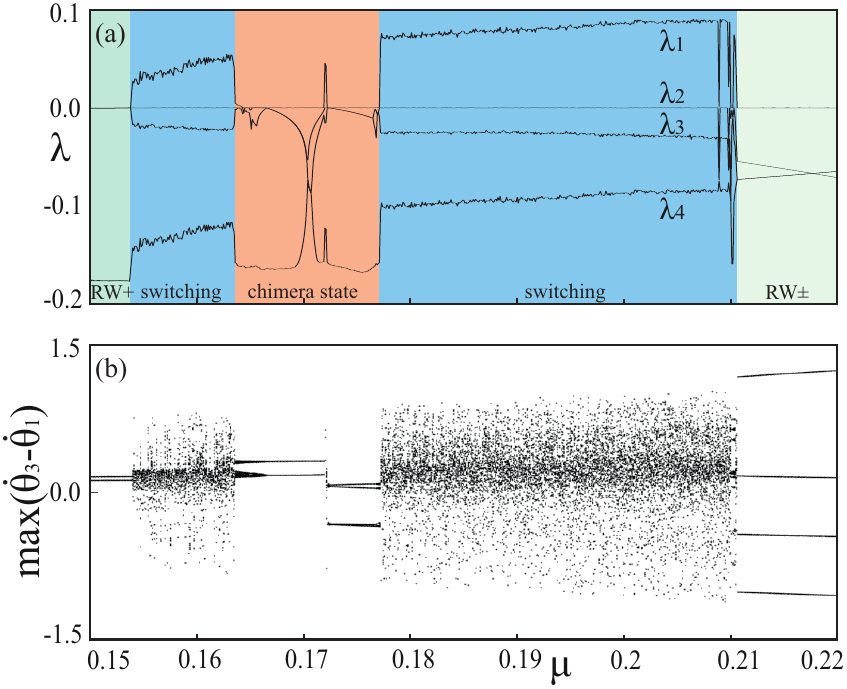}
	\caption{(color online) Lyapunov exponents (a) and bifurcation diagram for  reduced model (2) along the dashed vertical line at $\alpha=0.78$ in Fig. 1(b). Parameters $N=3, m=1.0$, and $\varepsilon=0.1$. }
	\label{fig3}
\end{figure} 

Results of direct numerical simulation of the model (1) in the two-parameter plane of the phase lag $\alpha$ and coupling strength $\mu$ are presented in Fig.1. Fig.1(a) reveals the appearance of numerous {\it chimera islands} scattered through the parameter plane. Chimeras in the islands are characterized by two phase-locked and one drifting oscillators, thus satisfying the condition of a weak chimera state \cite{ab2015}. The islands are surrounded in the parameter plane by traveling and chaotic switching waves as shown at enlargement in Fig.1(b). They arise in a saddle-node bifurcation, which is different from the standard all-to-all coupled Kuramoto model with inertia, where it is in a homoclinic bifurcation \cite{jar2018,smallest3_theor}.  

Generally, in Fig.1(a) we can distinguish  two {\it archipelagos} of the chimera islands: those to the left and to the right of the $\alpha=\pi/6$ vertical line. Visually, they look quite different, which can be caused by the fact that to the left of $\pi/6$ only one phase-locked state $O$ is stable (at small $\mu$), while two such such states, $O$ and $S$, stabilize to the right. With an increase of $\mu$, $O$ and $S$ lose their stability (at the curves $\mu=2\varepsilon^2/{\cos\alpha}$ and $0<\mu<2\varepsilon^2/{\cos(\alpha-2\pi/3)}$, respectively) producing two rotating waves $RW^{(-)}$ and  $RW^{(+)}$ in a supercritical Hopf bifurcation.  With more increase of $\mu$, the waves become chaotic giving a birth of the chimera islands inside.

In the blue region surrounding the four islands at $\alpha>\pi/6$ (see Fig.1(b)),  the system behavior represents a "stormy" chaotic switching between waves $RW^{(+)}$ and $RW^{(-)}$ rotating in opposite directions (illustrated in Fig.2(a,b)).   Stability regions for  $RW^{(+)}$ and $RW^{(-)}$ lie below the blue switching area, at some distance from the islands. The switching arises when both waves, $RW^{(+)}$ and $RW^{(-)}$,  lose stability one after another transforming into saddle states.  The behavior is then given by a chaotic attractor $A$ in the four-dimensional phase space of the reduced Eq.(2), representing a chaotic heteroclinic cycling between two saddle periodic orbits $P^{(+)}$ and $P^{(-)}$, images of  $RW^{(+)}$ and $RW^{(-)}$ (illustrated in Fig.2(c,d,g)).  Attractor $A$ should have in this case a double scroll shape, generally asymmetric, but obeying the symmetry at $\alpha=\pi/3$. Note also that in our simulations, we have not observed any other attractors types in the blue switching area;  this resembles the situation with well-known Lorenz attractor. 

It has also to be noted that, due to the axis symmetry with respect to $\alpha=\pi/3$,  chimera islands are identically equal to the left and to the right of this $\alpha$-value, and corresponding chimeras can be mapped into each other by a simple linear transformation specified above.

A typical scenario of the coherence-incoherence chimera transition is illustrated in Fig.2, where we fix $\alpha=0.78$ and increase the coupling strength $\mu$ along the vertical dashed line with triangles in Fig.1(b). First, in Fig.2(a), the system behavior is given by the rotating wave $RW^{(-)}$. It is born (at some  smaller $\mu$) out of the synchronous state $O$ and is rotating with negative velocity.  $RW^{(-)}$ co-exists with the splay state $S$ rotating in opposite direction. At slight increase of $\mu$, $RW^{(-)}$ losses its stability transforming into a saddle and, soon after, wave $RW^{(+)}$  with positive oscillator velocity grows from the splay state $S$  [Fig.2(b)].  With further  increase of $\mu$, in turn, wave $RW^{(+)}$ losses its stability (at $\mu\approx0.152$), transforming into a saddle state as well.  This is a bifurcation point for the coherence-incoherence transition in model (1): Beyond this parameter value, the behavior is developing in the form of chaotic homoclinic switching between the two saddle rotating waves  $RW^{(+)}$ and $RW^{(-)}$.  Fig.2(c) illustrates a typical switching solution for parameter $\mu$ values slightly above this bifurcation value: the solution spends more time close to wave $RW^{(+)}$, and only shortly visiting $RW^{(-)}$. At further increase of $\mu$, switching events between $RW^{(+)}$ and  $RW^{(-)}$ become more uniform [Fig.2(d)]. 

At $\mu\approx0.1634$, a chimera state arises terminating the switching in model (1). As it can be seen in Fig.3, first,  chimera behavior is quasi-periodic, but soon after it becomes periodic. Inside the chimera interval, however, there is a small gap around the value $\mu=0.172$, where chimera loses stability. The behavior is then developing in the form of chaotic heteroclinic switching between the three saddle chimera states (existing due to the permutational symmetry of the model (1)),  and between the two saddle rotating waves $RW^{(+)}$ and  $RW^{(-)}$  [Fig.2(f)]. The instability $\mu$-gap is very narrow at the chosen $\alpha$-value $0.78$, approximately it is  $[0.17293, 0.17223]$. It becomes, however, much wider at larger $\alpha$, which can be seen in the inset in Fig.1(b) showing a complicate microscopic structure  of the chimera region analyzed. The region shape may be associated with the so-called "shrimps", a typical as for stability regions in general dynamical systems [...]. 

Beyond the gap, the chimera states stabilize again obeying the same shape as before. They disappear at further increase of $\mu$, at about $\mu=0.1775$.  The system dynamics returns to the chaotic switching behavior. It is characterized now by more prolonged $RW^{(+)}$-intervals compared to $RW^{(-)}$ [Fig.2(g)]. For further increase of $\mu$,  a new rotating wave denoted $RW^{(\pm)}$ arises. In contrary to $RW^{(+)}$ and  $RW^{(-)}$,  it contains  both positive and negative rotations. (Note that the new arisen wave $RW^{(\pm)}$ co-exists, at some  parameter $\mu$-interval,  with the chaotic switching behavior, as it is also indicated in Fig.1(b)).   Soon after with further increase of $\mu$,  chaotic switching ceases to exist, and wave  $RW^{(\pm)}$ becomes the only system attractor.  
To finalize the description of the switching chimera transition in model (1), we present in Fig.3 respective graphs of the Lyapunov exponents and the bifurcations along the parameter $\mu$ interval considered.

  In our simulations, we have not observed any additional chimera islands in the parameter $\alpha>\pi/6$, except for the four shown in Fig.1(b).  On the other hand, as it can be seen in Fig.1(a), one big and plenty of small chimera islands are found scattered through the parameter plane at $\alpha<\pi/6$. In the present study, we do not go into details of the behavior in this part of the parameter plane, leaving this topic for future study.  The mechanism of the chimera phenomenon here, we assume, can be to some extent different from the described above, as only one regular wave,  $RW^{(-)}$ born out of synchronous state $O$, participates in generating the chaotic wave behavior as $\mu$ increases and $RW^{(-)}$ transforms to a saddle.  The chaotic attractor arising in this case (responsible for the wave behavior around the chimera islands) is then expected to obey a different, one-scroll shape. 
  
\section{COUPLED CENTURY: N=100 CASE.}
\begin{figure}
	\includegraphics[scale=1]{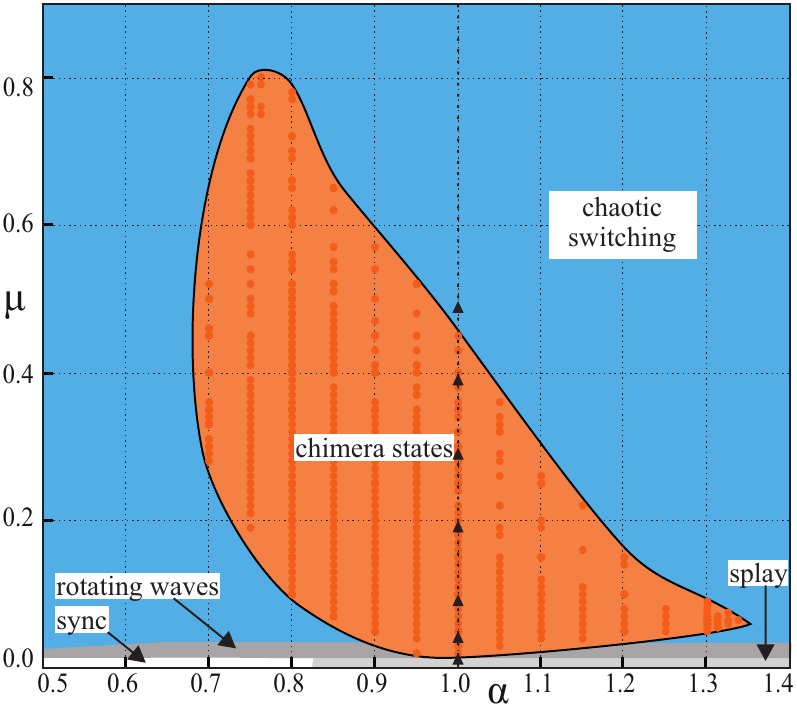}
	\caption{(color online) Region of chimera states for model (1) with $N=100$ and $P=40$ in the $(\alpha, \mu)$ parameter plane.  The red brick points inside the region indicate the parameter values in which the chimeras were obtained by direct numerical simulation with random initial conditions. In the blue area around, a sort of spatiotemporal chaotic behavior develops. At small values of the coupling parameter $\mu$ the behavior is regular in the form of fully synchronous (to the left) or splay (to the right) states, and periodic or quasiperiodic rotaing waves. Other parameters $m=1.0$ and $\varepsilon=0.1$.}
	\label{fig4}
\end{figure}

\begin{figure*}[ht]
	\begin{center}
		\includegraphics[scale=1]{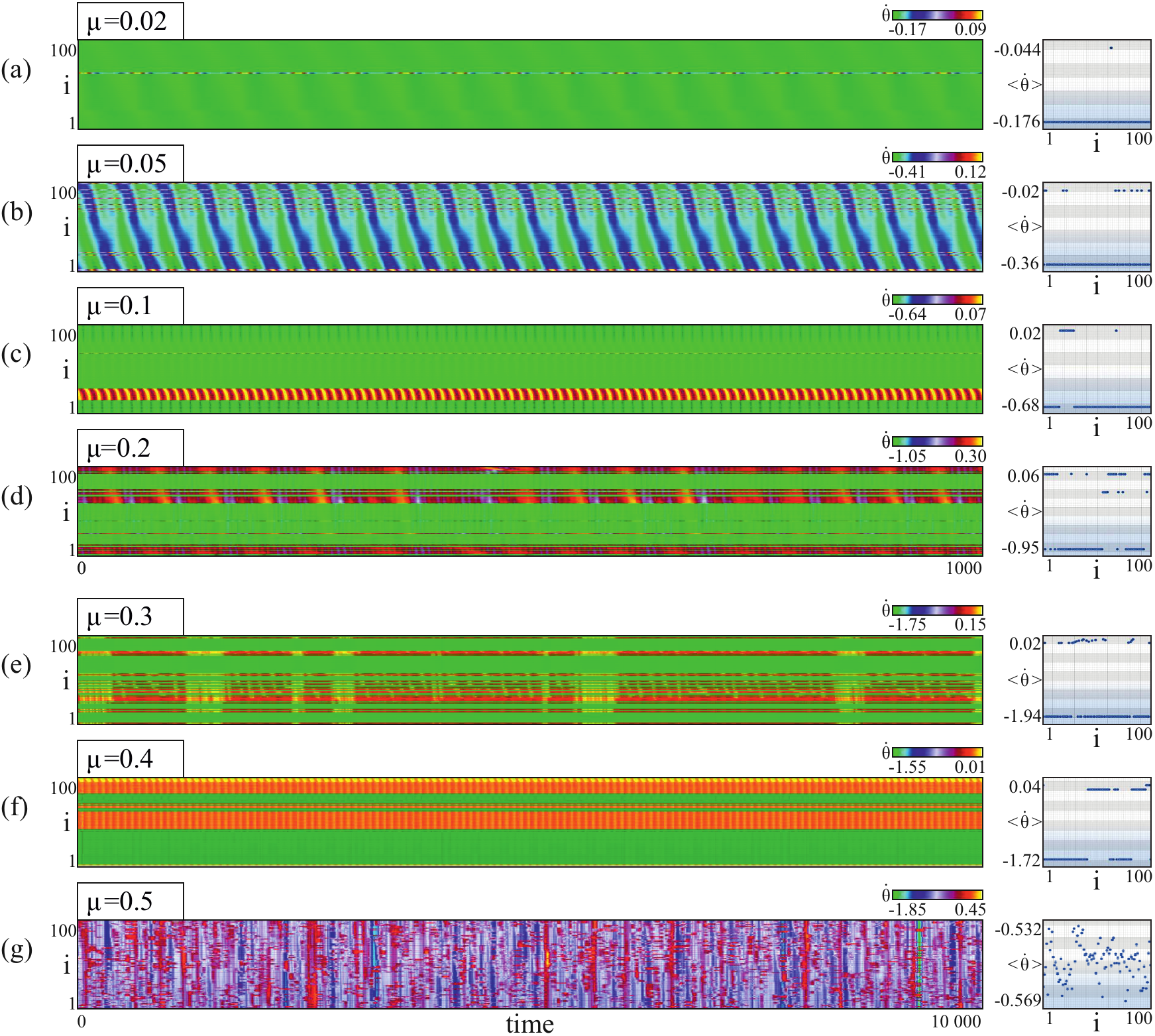}
		\caption{(color online) Chimera transition in model (1) with $N=100$ and $P=40$. Frequency-time plots (left panel) and average frequencies of individual oscillators (right panel) are shown for fixed phase lag $\alpha=1.0$  and coupling  $\mu$ along the dashed vertical line with triangles in Fig.4 (increasing from top to bottom: $\mu$=0.01, 0.05, 0.1, 0.2, 0.3, 0.4, 0.5, respectively).  (a) solitary state with a singe oscillator detached from the main synchronized cluster; (b) 11 detached oscillators scattered in one frequency cluster;  (c): 14 detached oscillators localized in one frequency cluster;  (d,f) 40 detached oscillators in two frequency clusters;  (e) chaotic chimera state with one fussy frequency cluster; (g) spatio-temporal behavior oabove the solitary region.  All simulations with random initial conditions. Other parameters $m=1.0$ and $\varepsilon=0.1$.} 			
		\label{fig5}
	\end{center}
\end{figure*}
To test if the chimera appearance is a universal phenomenon in directed networks, we have also investigated model (1) with $N=100$ and $P=40$. We find that there is only one, big chimera island in the parameter plane, which accumulates chimeras of various configurations, i.e. including a different number of the detached oscillators and different configurations, see Fig.4 and Fig.5.

 The chimera island in Fig.4 is surrounded by the chaotic switching area (shown in blue color). The chaotic waves "wash" the island from all sides, except a narrow strip from below (at small $\mu$),  where the dynamics is regular.  It is given by two phase-locked states, fully synchronous and splay states (at $\mu$-values of the order $0.01$), giving then a birth to periodic rotating waves (at $\mu$ of the order $0.02$) and becoming quasi-periodic at $\mu\approx0.03$. Note that the waves can co-exist with the solitary states.
 
  Typical solution behaviors through the chimera island are illustrated in Fig.5, where we fixed the phase lag $\alpha=1.0$ and increase the coupling strength $\mu$ along the vertical line with triangles in Fig.4  (all simulations are with random initial conditions).  First, in Fig.5(a), the chimera solution represents a typical {\it one-solitary state} \cite{jar2018} for $\mu=0.01$. Thus, a single oscillator is detached from the main synchronized cluster and start to rotate with a different average frequency, as shown in the  the right panel. At slight increase of $\mu$, the number of the frequency detached oscillators grows. We observe two possibility for the  distribution in the solitary oscillators in the ring:  they can scatter through the space in a visually random way [Fig.5(b)] or concentrate at some parts of the ring, mainly in one or two interval [Fig.5(c,f)].  With further increase of $\mu$, two different chimera states are generally developed, with regular and chaotic behaviors, the examples are illustrated  in Fig.5(d,f,e), respectively. In the regular case, the so-called {\it quasi-periodic chimera} can arise with two frequency clusters for the detached oscillators [Fig.5(d,f), right panel]). In the chaotic case, the detached oscillators are chaotically distributed forming a fussy cluster [Fig.5(e)]. Eventually, above the chimera island, the behavior represents a sort of the spatio-temporal chaos with a fussy intervals of the both side collective rotations and the individual oscillators trying unsuccessfully to escape.

 \section{ DISCUSSION AND CONCLUSION}
  
  In our study, we have analyzed the unidirectionally coupled model (1) with  different values of the system size $N$ and the coupling radius $P<N/2$. We confirmed the chimera appearance for the following pairs of small $N$ and $P$:  $N=5, P=2$;  $N=7, P=3$; $N=9, P=3$ and $4$, and we where not able to obtain any chimera states for coupling radius $P$=1 (except of the $N=3$ case reported above).  Interestingly, in each of the cases, the chimera regions are quite large, compared to the $N=3$ case. 
  Regarding the $N=100$ case, our simulations approve that the chimera island shown in Fig.4 for $N=100$ and $P=40$ decreases gradually with decreasing of $P$. The minimal island observed in our simulations was at $P=20$, and we were not able to catch any chimera state at $P=15$. 
  
In conclusion, we have identified the chimera states in a network of $N$   unidirectionally coupled oscillators on a ring, and have described the mechanism of their appearance for $N=3$ and $N=100$. The results are both surprising and counter-intuitive for networks with unidirectional coupling, e.g. in feed-back delayed \cite{delay1,delay2,delay3} and feed-forward \cite{park2010} systems, where a reductions to the ring of oscillators  with unidirectional coupling is possible. For the time-delayed systems, the chimeras can be assiciated then with the dissipative solitons, see \cite{bru2018, sem2018}.   We expect that the amazing chimera behavior discovered can indicate a common, probably universal phenomenon in the directed networks of very different nature having a lot of applications in practice.

\textbf{Acknowledgment}
 This work has been supported by the National Science Centre, Poland, OPUS Programme (Project No 2018/29/B/ST8/00457).

\textbf{Data availability}: The data that support the findings of this study are available from the corresponding author upon reasonable request.


\begin{thebibliography}{70}
	
\bibitem{pa2015}  M. J. Panaggio and D. M. Abrams, Nonlinearity 28, R67 (2015). 

\bibitem{s2016} E.~Sch\"{o}ll, Eur. Phys. J. Special Topics 225, 891-919 (2016).

\bibitem{omel2019} O.~Omel'chenko and E.~Knobloch, New J. Phys. 21, 093034 (2019).

\bibitem{anna2020} A.~Zakharova {\it Chimera Patterns in Networks: Interplay between Dynamics, Structure, Noise, and Delay}. Springer Nature (2020).

\bibitem{chimeras}  F. Parastesh, S. Jafari, H. Azarnoush, Z. Shahriari, Z. Wang, S. Boccaletti, and M. Perc, Phys. Rep. 898, 1 (2021).
	
\bibitem{network} J. Hizanidis, N.E. Kouvaris, G. Zamora-Lopez, A. Diaz-Guilera, and C. Antopoulos, Sci.Rep. 6, 19845 (2016).
	
\bibitem{birds} N.C. Rattenborg, C.J. Amlaner, and S.L. Lima, Neurosci. Biobehav. Rev. 24, 817 (2000).
	
\bibitem{epilepsy} A. Rothkegel and K. Lehnertz, New J. Phys. 16,0055006 (2014).
	
\bibitem{neuro} Z. Wang and Z. Liu, Front. Physiol. 11:724 (2020)

\bibitem{grid} A.E. Motter, S.A. Myers, M. Angel and T. Nishikawa, Nat. Phys., 9, 191 (2013).

\bibitem{pshmr2014} L. M. Pecora F. Sorrentino, A. Hagerstrom, T. Murphy, and R. Roy,  Nat. Commun. 13 4079 (2014).

\bibitem{squids} N. Lazarides and G. Tsironis, Phys. Reports 752 1-67 (2018).

\bibitem{belykh2017} I. Belykh, R. Jeter, and V. Belykh, Sci. Adv. 3, e1701512 (2017).

\bibitem{delayed} J. Hart, L. Larger, T. Murphy, and R. Roy, Phil. Trans. R. Soc. A 377, 20180123 (2019).

\bibitem{powergrid} F. Hellmann, P. Schultz, P. Jaros, R. Levchenko, T. Kapitaniak, and J. Kurths, Nat. Commun. 11, 592 (2020).

\bibitem{social} J.C. Gonzales-Avella, M.G. Cosenza and M.S. Miguel, Physica (Amsterdam) 399A, 24 (2014).

\bibitem{pik2021} A.~Pikovsky, Math.Model.Nat.Phenom. 16, 15 (2021).  
		
\bibitem{ref2} 	Y. Kuramoto, D. Battogtokh, Nonlinear Phenom. Complex Systems 5, 380-385 (2002)
	
\bibitem{ref3} 	D. M. Abrams, S. H. Strogatz, Phys. Rev. Lett. 93, 174102 (2004). 
	
\bibitem{drift2010} O.Omelchenko, M. Wolfrum, and Yu. Maistrenko, Phys.Rev.E 81, 065201(R) (2010).

\bibitem{asym2013} V. Dziubak, V. Maistrenko, Yu. Maistrenko, and M. Timme (2013) unpublished.

\bibitem{xie2014}J. Xie, E. Knobloch and H. C. Kao, Phys. Rev. E 90, 022919 (2014).
	
\bibitem{bick2015} C. Bick and E. A. Martens, New J. Phys. 17 033030, (2015).
	
\bibitem{kkwcm2014} T. Kapitaniak, P. Kuzma,  J. Wojewoda, K. Czolczynski, Yu. Maistrenko, Sci. Rep. 4, 6379 (2014). 

\bibitem{jar2015} P. Jaros, Y. Maistrenko, and T. Kapitaniak, Phys. Rev. E 91, 022907
(2015).

\bibitem{jar2018} P. Jaros,  S. Brezetsky,  R. Levchenko,  D. Dudkowski,  T. Kapitaniak,  Yu. Maistrenko, Chaos 28, 011103 (2018).

\bibitem{jar2021}  S. Brezetsky, P. Jaros, R. Levchenko,  T. Kapitaniak,  Yu. Maistrenko, Phys. Rev. E 103, L050204 (2021).

\bibitem{ab2015} P. Ashwin and O.Burylko, Chaos 25, 013106 (2015).

\bibitem{adaptive}  R.~Berner, A.~Polanska, E.~Sch\"{o}ll, S.~Yanchuk, 
Eur. Phys. J. Special Topics 229, 2183 (2020).

\bibitem{anna2021}	L. Sch\"{u}len, D. Janzen, E. Medeiros, and A. Zakharova, Chaos, Solitons and Fractals 145, 110670 (2021).

\bibitem{kruk2020} N.~Kruk, Yu.~Maistrenko, and H.~Koeppl, Chaos 30, 111104 (2020).


\bibitem{onbt2014} S. Olmi, A. Navas, S. Boccaletti, and A. Torcini, Phys. Rev. E 90, 042905; S. Olmi, E. A. Martens, S. Thutupalli, and A. Torcini, Phys. Rev. E 92,
030901(R) (2015); S. Olmi, Chaos 25, 123125 (2015).

\bibitem{smallest3_theor} Y. Maistrenko, S. Brezetsky, P. Jaros, R. Levchenko, and T. Kapitaniak, Phys. Rev. E 95, 010203(R) (2017).
	
\bibitem{delay1} V. Klinshov, D. Shchapin, S. Yanchuk, M. Wolfrum, O. D'Huys, V. Nekorkin
Phys. Rev. E 96, 042217 (2017).
	
\bibitem{delay2} J.D. Hart, L. Larger, T.E. Murphy and R. Roy, Phil. Trans. R. Soc. A.377 (2019).

\bibitem{delay3} V. Semenov, A. Zakharova, Y. Maistrenko, and E. Sch\"{o}ll, Europhys. Lett.
115, 10005 (2016).
	
\bibitem{park2010} Y.-S. Park et al., IEICE Trans. Electron E93-C  9, 1467-1470 (2010).

	
 \bibitem{bru2018} D. Brunner , B. Penkovsky , R. Levchenko, E. Sch\"{o}ll, L. Larger, andY. Maistrenko, Chaos 28, 103106 (2018).

\bibitem{sem2018} V. Semenov and Yu. Maistrenko, Chaos 28, 101103 (2018).
	
	
		
	
	
	
	
	
	
	
	
	
	
	















	
\end{thebibliography}
\end{document}